\documentclass[final,5p,times,twocolumn]{elsarticle}
\usepackage{amssymb,amsmath}
\usepackage{hyperref}
\biboptions{sort&compress}
\journal{Physics Letters B}

\begin{document}
\begin{frontmatter}


\title{Comparison of fission and quasi-fission modes}

% Force line breaks with \\
\author[add1]{C. Simenel\fnref{label1}}
\fntext[label1]{Corresponding author}
\ead{cedric.simenel@anu.edu.au}
\author[add1]{P. McGlynn}
\author[add2]{A.S. Umar}
\author[add3]{K. Godbey}
\address[add1]{Department of Fundamental and Theoretical Physics and Department of Nuclear Physics and Accelerator Sciences, Research School of Physics,\\ The Australian National University, Canberra ACT  2601, Australia}
\address[add2]{Department of Physics and Astronomy, Vanderbilt University, Nashville, Tennessee 37235, USA}
\address[add3]{Facility for Rare Isotope Beams, Michigan State University, East Lansing, Michigan 48824, USA}
%------------------------------------------------------------------------------

\begin{abstract}
Quantum shell effects are known to affect the formation of fragments in nuclear fission. Shell effects also affect quasi-fission reactions occurring in heavy-ion collisions. Systematic time-dependent Hartree-Fock simulations of $^{50}$Ca$+^{176}$Yb collisions show that the mass equilibration between the fragments in quasi-fission is stopped when they reach similar properties to those in the asymmetric fission mode of the $^{226}$Th compound nucleus.  Similar shell effects are then expected to determine the final repartition of nucleons between the nascent fragments in both mechanisms. Future experimental studies that could test these observations are discussed.
\end{abstract}
\begin{keyword}
Quasifission \sep fission \sep TDHF



\end{keyword}

\end{frontmatter}


\section{Introduction}
\label{intro}

Nuclear fission and quasi-fission are {\it a priori} very different reaction mechanisms. 
On the one hand, fission occurs when a heavy nucleus splits into two (or more) fragments. 
The fissioning nucleus can be initially in its ground-state, as in spontaneous fission, or in an excited state, as in neutron-induced fission or in fission following fusion of two heavy ions. 
In the latter two cases, a  compound nucleus is  formed with equilibrated internal degrees of freedom in such a way that 
the fission process only depends on its excitation energy and angular momentum. 
On the other hand, quasi-fission is an out-of-equilibrium mechanism occurring when two heavy collision partners transfer a significant amount of nucleons through mass equilibration, 
before separating in fission-like fragments without the intermediate formation of a compound nucleus \cite{toke1985} (see \cite{hinde2021} for a recent experimental review). 

Nevertheless, both processes also exhibit some similarities. 
For instance, the total kinetic energy of the fragments is well approximated by the Viola systematics \cite{viola1985,hinde1987}, indicating a slow, damped relative motion of the fragments. 
In addition, the timescale for quasi-fission reactions \cite{toke1985,durietz2013,simenel2020} is of the same order as the minimum average timescale for the evolution from the compound system to the formation of the final fragments which is about $20-50$~zs  (1~zeptosecond  $=1$~zs$=10^{-21}$~s) \cite{hinde1992}. 
Another similarity is that both reaction mechanisms are impacted by quantum shell effects. 
In fission, shell effects are able to drive the system away from mass symmetric fission, while in quasi-fission, they are expected to stop the mass equilibration process.

The purpose of this work is to compare such quasi-fission and fission modes that are driven by shell effects.
Although several shell effects are expected to occur in the compound system on its way to fission~\cite{bernard2021}, 
our focus is on  those in the nascent fragments that are responsible for the final repartition of protons and neutrons between the fragments. 
Neutron-induced actinide  fission \cite{brown2018} and fission of neutron deficient actinides \cite{schmidt2000,bockstiegel2008,chatillon2019} reveal the presence of an asymmetric fission mode producing heavy fragments with $Z\simeq54$ protons. Octupole (pear shape) deformed shell effects at $Z=52$ and $56$ \cite{scamps2018} have been invoked to explain the constancy of the heavy fragment charge distribution centroid.
Spherical shell effects in the $^{132}$Sn region with magic numbers $Z=50$ and $N=82$ are also known to induce a symmetric fission mode in neutron-rich fermium isotopes \cite{hulet1986}.
In addition, other deformed shell effects are being investigated in near and sub-lead region \cite{scamps2019,mahata2020,prasad2020,swinton2020b} to explain asymmetric fission observed in this region \cite{andreyev2010,prasad2015}. Furthermore, spherical shell effects in $^{208}$Pb are predicted to induce a {\it super-asymmetric} mode in some superheavy nuclei (SHN) \cite{poenaru2011,warda2018,matheson2019,poenaru2018,zhang2018,santhosh2018,ishizuka2020}. 
However, no experimental confirmation of the latter exist so far due to the difficulty of creating superheavy compound nuclei \cite{vardaci2019}.

Shell effects have also been invoked to explain quasi-fission fragment mass distributions \cite{itkis2004,nishio2008,kozulin2010,nishio2012,kozulin2014,wakhle2014,itkis2015,morjean2017,hinde2018,kozulin2021}. 
In particular mass equilibration is often stopped, in reactions forming SHN, when a heavy fragment in the doubly magic $^{208}$Pb region is produced, even at energy well above the Coulomb barrier \cite{wakhle2014,oberacker2014,umar2016}.
The first experimental confirmation of this effect was only recently obtained through measurement of  X-rays from the quasi-fission fragments indicating an excess of fragments with the proton magic number $Z=82$ \cite{morjean2017}.
Although this observation of a quasi-fission mode produced by shell effects could potentially be associated to the predicted super-asymmetric mode in SHN fission, there has been so far no observation (either experimentally or in numerical simulations) of quasi-fission modes that could be identified to known fission modes. 

Our purpose is then to investigate quasi-fission modes in a heavy-ion reaction which, in the case of fusion, would produce a compound nucleus with known fission modes. 
We choose the $^{50}$Ca$+^{176}$Yb reaction at an energy of $13\%$ above the Coulomb barrier.
The choice for this reaction is motivated by the fact that its compound nucleus, $^{226}$Th, is  known experimentally to have two fission modes, one symmetric and one asymmetric, both with similar  yields \cite{schmidt2000,bockstiegel2008,chatillon2019}. 
Our goal is then to investigate if quasi-fission is able to populate one or both of these fission modes. 
Our theoretical modelling is based on the Hartree-Fock (HF) self-consistent mean-field theory with a Skyrme energy density functional (EDF), which is known to account properly for shell effects in nuclear systems \cite{bender2003}. 

\section{Results}

Our approach is based on three steps. First, we study the fission modes in $^{226}$Th. 
Although theoretical modelling of fission is still an ongoing challenge~\cite{bender2020}, 
microscopic approaches are commonly used to investigate fission modes. 
Here, we construct a potential energy surface (PES) with the constrained-HF method with BCS pairing correlations.
This PES is used to confirm that our choice of EDF leads to two fission modes, associated with a symmetric and an asymmetric valleys. 
 %, and compare them with experimental data \cite{bockstiegel2008}.
Second, we perform a systematic study of $^{50}$Ca$+^{176}$Yb collisions with the time-dependent Hartree-Fock (TDHF) theory, searching for quasi-fission trajectories. 
Finally, we search for potential quasi-fission modes and compare them with the fission ones. 
This theoretical approach is motivated by the fact that the same EDF is used to describe both nuclear structure and reaction dynamics, and by the now well established applicability of TDHF to study quasi-fission in a broad range of systems \cite{simenel2020,wakhle2014,oberacker2014,umar2016,umar2015a,hammerton2015,sekizawa2016,guo2018d,zheng2018,godbey2019} (see \cite{simenel2012,simenel2018,sekizawa2019,stevenson2019} for recent reviews of TDHF applications to  heavy-ion reactions).
In particular, the approach has no free parameters as its only phenomenological input is the Skyrme EDF whose parameters are usually determined from properties of some nuclei and of infinite nuclear matter.  We chose the SLy4$d$ parametrisation \cite{kim1997} which can be used  in static calculations as well as to simulate heavy-ion collisions.
%See \cite{supplemental} for a description of the constrained-HF and TDHF calculations. 

\begin{figure}[!htb]
\centerline{\includegraphics[width=8.6cm]{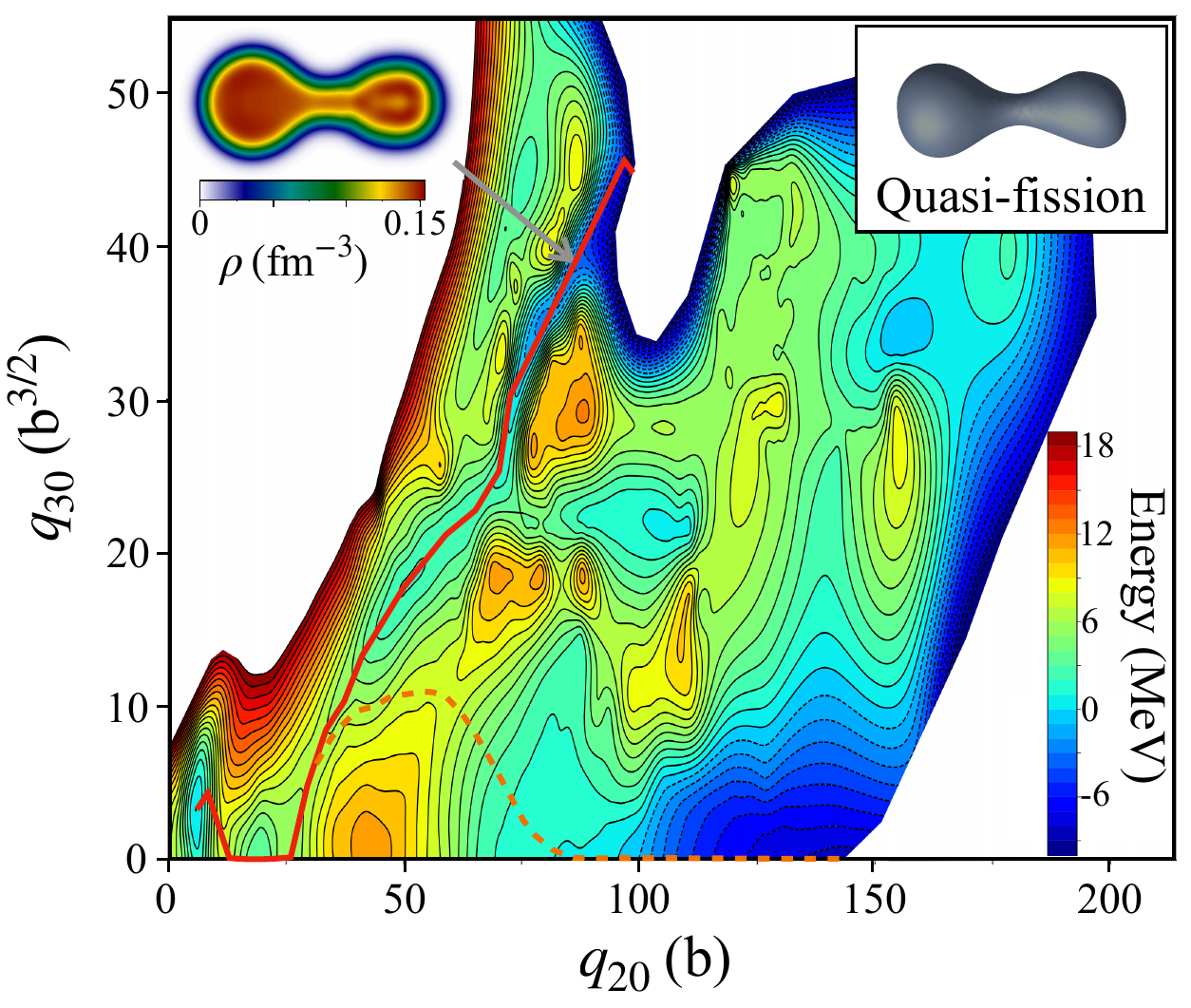}}
\caption{Potential energy surface of $^{226}$Th obtained from the constrained Hartree-Fock method with BCS correlations using the SkyAx solver with quadrupole and octupole steps $\Delta q_{20}=1.44$~b and $\Delta q_{30}=1$~b$^{3/2}$, respectively. 
The red solid line shows the fission path obtained by leaving the octupole moment free. The dashed line, showing a possible path towards the symmetric valley, is to guide the eye. A density distribution for the system along the asymmetric fission path is shown for near-scission deformation indicated by the arrow. An isodensity surface (grey) at half the saturation density ($\rho_0/2=0.08$~fm$^{-3}$) is shown for a similar near-scission configuration (represented by star symbols in Fig.~\ref{fig:traj}) obtained from a TDHF quasi-fission trajectory. }
\label{fig:PES}
\end{figure}

To investigate the fission modes of $^{226}$Th with the SLy4$d$ Skyrme functional, a potential energy surface is constructed from mean-field  solutions under constraints on the quadrupole moment 
 $$q_{20}=\sqrt{\frac{5}{16\pi}}\int d^3r \,\rho(\mathbf{r}) (2z^2-x^2-y^2),$$
where $\rho(\mathbf{r})$ is the density of nucleons, 
 fixing the elongation of the system, and the octupole moment 
  $$q_{30}=\sqrt{\frac{7}{16\pi}}\int d^3r \,\rho(\mathbf{r}) [2z^3-3z(x^2+y^2)]$$
 fixing its asymmetry. 
For this purpose we use the SkyAx code which solves the constrained HF equations with BCS pairing correlations and axial symmetry \cite{reinhard2021}. 
The resulting PES in Fig.~\ref{fig:PES}  confirms the existence of two fission valleys. 
The first valley is mass asymmetric.  It is present in many actinides and usually lead to $Z_H\simeq52-56$ protons in the final heavy fragments \cite{andreyev2017,schmidt2018}.  
This is the valley explored by the system if the octupole moment is not constrained (solid line). We also see that the system may return to symmetric shapes ($q_{30}=0$) for little additional cost in energy (dashed line), leading to symmetric elongated fragments. 
Indeed, the difference in energies between the saddle point to  return to the symmetric valley (overcome by the dashed line) and the first saddle point is only 1.2 MeV.
We therefore expect both symmetric and asymmetric fission modes to occur with similar probabilities. 
These results are in good agreement with  theoretical predictions using other EDF (see, e.g., \cite{dubray2008,tao2017,bernard2020}), as well as with experimental observations indicating similar yields for both modes at low excitation energy~\cite{bockstiegel2008,chatillon2019}. % (see Fig.~\ref{fig:Exp}a). 

\begin{figure}[!htb]
\centerline{\includegraphics[width=8.6cm]{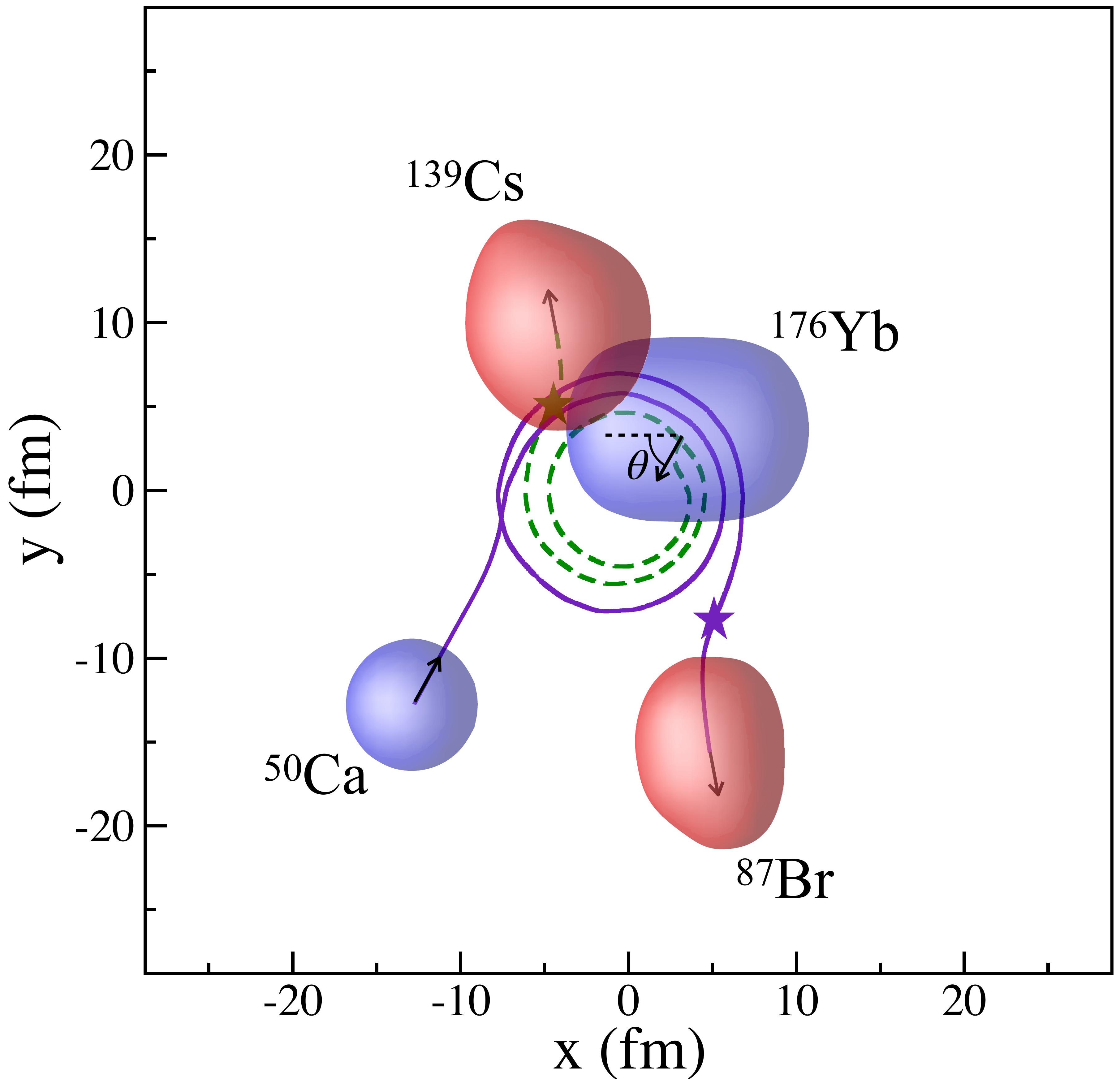}}
\caption{Example of  TDHF calculation of $^{50}_{20}$Ca$+^{176}_{\,\,\,\,70}$Yb with an orbital angular momentum $L=82\hbar$ leading to quasi-fission fragments $^{87}_{35}$Br$+_{\,\,\,\,55}^{139}$Cs. The surfaces represent the initial (blue) and final (red) isodensities at half the saturation density  $\rho_0/2=0.08$~fm$^{-3}$. The solid and dashed lines represent the evolution of the centres of masses of the light and heavy fragments, respectively. The star symbols indicate the position on the trajectory used to represent the isodensity in Fig.~\ref{fig:PES}.
 The $x$ and $y$ scales   correspond to the full numerical box.}
\label{fig:traj}
\end{figure}

%We now search for quasi-fission trajectories in systematic TDHF calculations of $^{50}$Ca$+^{176}$Yb collisions.
Our goal is now to investigate quasi-fission modes in $^{50}$Ca$+^{176}$Yb collisions which, in the case of fusion, would form the $^{226}$Th compound nucleus. 
Quasi-fission is known to rapidly increase with the charge product $Z_1Z_2$ of the reactants \cite{hinde2021}.
Although experimental signatures of quasi-fission have been found in systems with $Z_1Z_2$ as small as 736 in the $^{16}$O$+^{238}$U reaction \cite{hinde1996}, the lightest system in which quasi-fission reactions have been observed in TDHF calculations is $^{50}$Cr$+^{180}$W ($Z_1Z_2=1776$) \cite{hammerton2015}.
As TDHF predicts the most likely outcome for a given initial configuration, only a small range of orbital angular momenta $L$ (or, equivalently, impact parameters) might lead to TDHF trajectories with quasi-fission characteristics in the $^{50}$Ca$+^{176}$Yb system as it has a relatively small charge product $Z_1Z_2=1400$. 

In this work, the TDHF3D code is used with  a plane of symmetry (the $z=0$ reaction plane) \cite{kim1997}.
BCS correlations are included in the initial static calculations to 
avoid spurious deformations in open shell nuclei. These correlations
are then treated with the frozen occupation approximation in the time evolution.
While the $^{50}$Ca mean-field ground-state is found to be spherical, the $^{176}$Yb ground-state is obtained with a prolate deformation $\beta_2\approx0.33$ and thus its orientation is expected to impact the reaction mechanism~\cite{godbey2020b}. 
The centre of mass energy of the reaction is $E_{c.m.}=172$~MeV, corresponding to approximately $13\%$ above the Coulomb barrier $V_B\simeq151.8$~MeV according to the systematics of Swiatecki  {\it et al.} \cite{swiatecki2005}. 
This energy is large enough to ensure that all initial orientations of the prolately deformed $^{176}$Yb may lead to contact between the collision partners and then potentially contribute to quasi-fission~\cite{hinde1996}. 
The initial distance between the centres of mass of the collision partners is 22.6~fm.
As the angle of emission of the fragments is unknown prior to a calculation, large Cartesian grids of $72\times72\times(28/2)\times\Delta x^3$ with a mesh size $\Delta x=0.8$~fm are used to allow for a full description of the exit channel with well separated final fragments. 
We performed 40 TDHF calculations with four initial orientations of $^{176}$Yb deformation axis (forming an angle of 0, 45, 90, and 135 degrees with respect to the axis joining the initial centres of mass), and with  a $\Delta L=2\hbar$ step in orbital angular momentum, for a total of 14,000~CPU hours on Intel Xeon Scalable `Cascade Lake' processors. 
The results are compiled in Supplemental Material Table~\ref{tab:results}.

An example of  resulting quasi-fission  trajectory is shown in Fig.~\ref{fig:traj} for an  orbital angular momentum $L=82\hbar$ and an  orientation of $\theta\simeq62.0$~degrees between the $^{176}$Yb initial velocity vector and its deformation axis. 
The position of the fragments is obtained at each time by computing the centres of mass of the density distributions on each side of the neck. 
The resulting trajectories are represented by the solid and dashed lines in Fig.~\ref{fig:traj}. 
We see that the system undergoes more than a full rotation, during which about 37 nucleons (in average)  are transferred from the heavy fragment to the light one. 
The total contact time, defined as the time during which the neck density exceeds half saturation density $\rho_0/2\simeq0.08$~fm$^{-3}$, is $\tau\simeq22.9$~zs for this collision. 
This contact time and the large amount of mass transfer between reactants are typical of quasi-fission reactions~\cite{toke1985,durietz2013}.

\begin{figure}[!htb]
\centerline{\includegraphics[width=8.6cm]{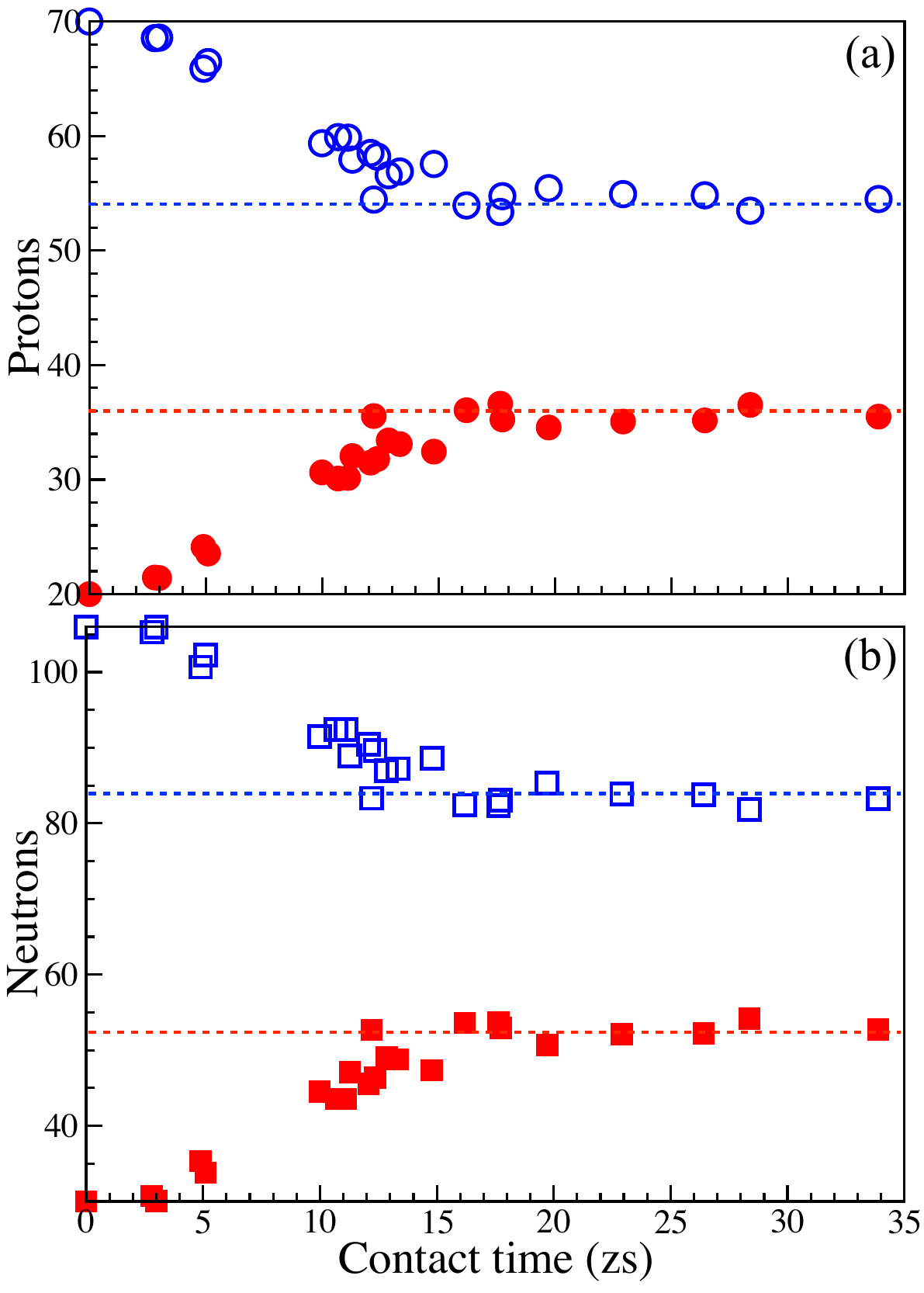}}
\caption{(a) Proton (circles) and  (b) neutron (squares) numbers in the heavy (open symbols) and light (filled symbols) fragments as a function of contact time. The dashed lines represent possible asymptotic values at $Z_H\simeq54$, $Z_L\simeq36$, $N_H\simeq84$, and $N_L\simeq52$.}
\label{fig:time}
\end{figure}

Quasi-fission trajectories were searched for up to approximately 30~zs contact times. 
Although a slow quasi-fission component  with longer contact times is observed experimentally \cite{hinde2021}, this upper limit is of the order of the longest quasi-fission times observed in TDHF calculations \cite{simenel2020}.
We therefore consider that the system has fused when the contact time reaches $\tau\sim31$~zs  (unless an increase of elongation indicates a likely quasi-fission at a later time, in which case the calculations is run up to $\tau\sim35$~zs), which occurs essentially below  critical angular momentum $L_{c}$  that depends on the orientation of the target.
Collisions that lead to contact with the  side of $^{176}$Yb are found to have smaller critical value $L_c\sim60\hbar$, while collisions with its tip lead to $L_c\sim 80\hbar$. 
At large $L$, only few nucleons are exchanged in quasi-elastic collisions, occurring at $L_q\sim68\hbar$ ($\sim96\hbar$)  in collisions with the side (tip) of $^{176}$Yb.
For a given orientation, quasi-fission is obtained  for $L_c \lesssim L \lesssim L_q$. 
(Note that in few cases we observe quasi-fission for $L<L_c$, see Supplemental Material Tab.~\ref{tab:results}.)  

The number of protons and neutrons in the outgoing fragments are plotted in Fig.~\ref{fig:time} as a function of the contact time. 
A correlation is observed at short contact times ($\tau\lesssim13$~zs) where nucleons are transferred from the heavy to the light fragment.
At longer contact times, however, this correlation is lost, with constant numbers of protons and neutrons $Z_L\simeq36$, $N_L\simeq52$, $Z_H\simeq54$ and $N_H\simeq84$ in the light and heavy fragments, respectively, indicating a stop of the mass equilibration process.
Interestingly, this occurs when the fragments have reached the same numbers of neutrons and protons as the asymmetric fission fragments of $^{226}$Th, which is a first indication that these fission and quasi-fission modes are similar. 

\begin{figure}[!htb]
\centerline{\includegraphics[width=8.6cm]{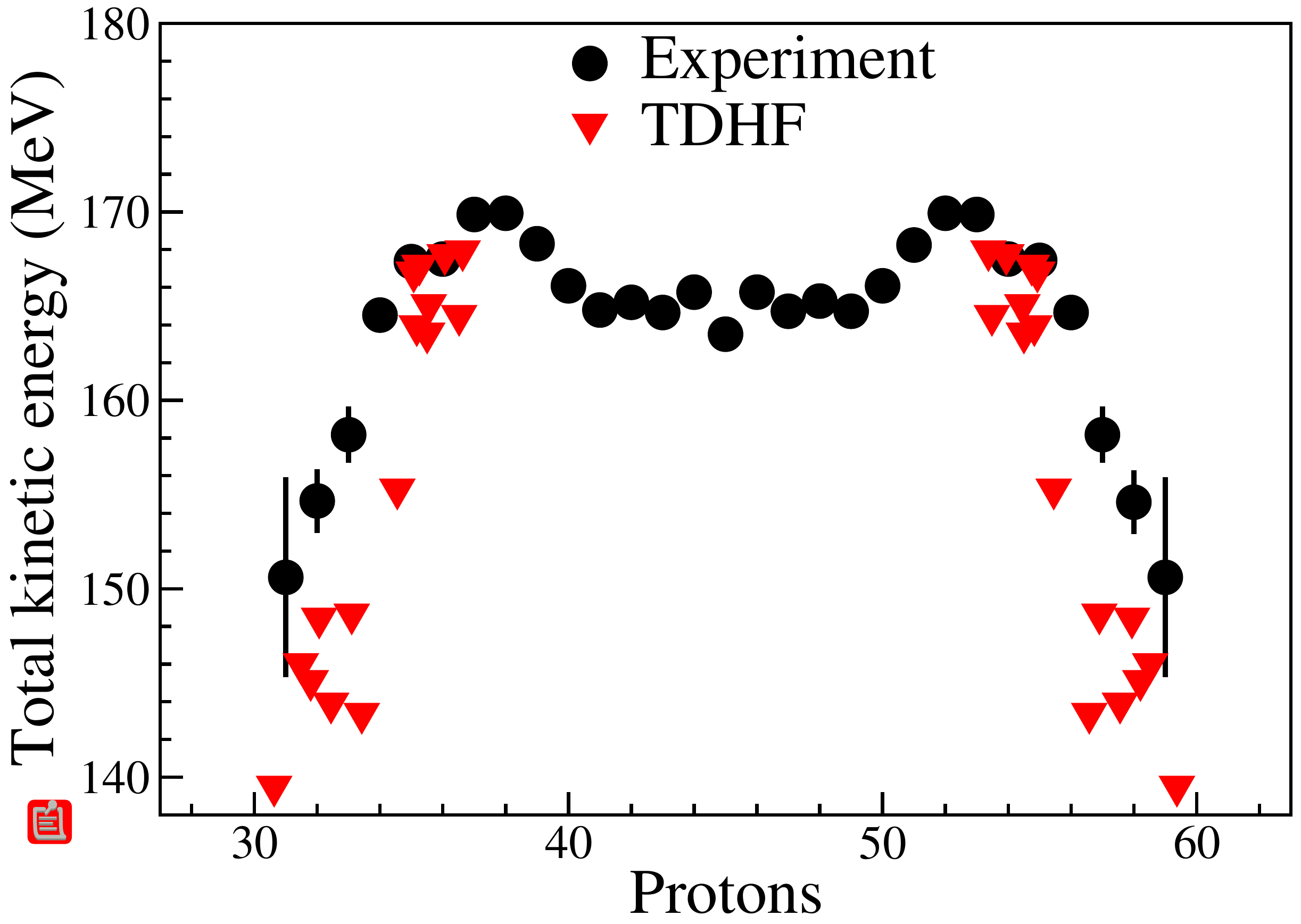}}
\caption{%(a) Experimental charge fragment distribution in low  energy fission of $^{226}$Th (circles) \cite{bockstiegel2008} and TDHF prediction of quasi-fission fragment production cross-sections in $^{50}$Ca$+^{176}$Yb (per bin of $\Delta Z=2$). (b) 
Experimental average TKE of $^{226}$Th fission fragments (circles) from Ref. \cite{bockstiegel2008} and TDHF predictions of quasi-fission fragments TKE (triangles).}
\label{fig:Exp}
\end{figure}

As most of the shell effects fixing the final asymmetry in nascent fission fragments are deformed shell effects, it is important to compare the shape of fragments as well as the number of protons and neutrons. 
Experimentally, the shape of the system at scission is inferred indirectly through the total kinetic energy (TKE) of the fragments. 
Indeed, for a similar mass and charge partition, higher TKE are larger for more compact fragments, while elongated fragments lead to lower TKE. 
Figure~\ref{fig:Exp} provides a comparison between experimental TKE of fission fragments and TKE of quasi-fission fragments obtained from TDHF by summing kinetic and Coulomb energy between the fragments (see, e.g., \cite{simenel2014a}). 
We see that, for the fragments which have reached the $Z_L/Z_H\approx36/54$ partition, the TKE are similar in both processes, indicating similar shapes of the systems at scission. 
However, for more asymmetric splits, quasi-fission leads to smaller TKE than fission which could be attributed to differences in the dynamics.  
Indeed, larger asymmetries in quasi-fission are obtained for the most peripheral collisions (larger $L$) inducing shapes which can significantly differ to those in fission. 
The fact that quasi-fission fragments with $Z_L/Z_H\approx36/54$ partition have a similar shape than those produced in the  asymmetric fission mode of $^{226}$Th is further supported by a comparison of the densities near scission in Fig.~\ref{fig:PES}.

The observation that the mass equilibration is stopped when the fragments have reached  proton and neutron numbers, as well as shapes, that are similar to those in the asymmetric fission mode, is an indication that both modes should have the same origin in terms of shell effects. 
In particular, the octupole deformed shell effects at  $Z_H\simeq52-56$, that were invoked as a mechanism allowing the heavy fragment to acquire pear shapes for a small cost (if any) in energy \cite{scamps2018}, could also be responsible for stopping the mass equilibration process in quasi-fission. 
 %In fact, it seems to be stopped at slightly larger asymmetry than in fission, as we did not observe quasi-fission heavy fragments with less than 53 protons, while in $^{226}$Th fission, significant yields are found down to $Z_H=52$ in the asymmetric mode \cite{bockstiegel2008,chatillon2019}.
%This could be due to the fact that, in the asymmetric fission mode, the system increases its asymmetry, while in quasi-fission it is decreased, and thus the systems could explore slightly different configurations in a relatively wide valley.

\section{Conclusions}
%As a result, for the system studied here, quasi-fission reactions could be used to explore the asymmetric fission valley. 
Like fission, quasifission is affected by shell effects through valleys 
in the potential energy surfaces \cite{zagrebaev2005,zagrebaev2007}. Here, we have 
shown that quasifission may populate the asymmetric fission mode of $^{226}$Th.
However, the symmetric fission mode, which has similar yields as the asymmetric one in $^{226}$Th, is not observed in our TDHF simulations of quasi-fission. 
A possible explanation is that the  shell effects responsible for the asymmetric fission mode are strong enough to stop mass equilibration in every quasi-fission trajectory. 
Alternatively, longer contact times may be needed for quasi-fission trajectories leading to symmetric fragments. In that case, beyond mean-field fluctuations and correlations that build up over time may be required. 
It would be interesting to investigate this system with the stochastic mean-field approach which, in addition to incorporating such fluctuations as demonstrated in the case of fission \cite{tanimura2017}, might allow the system to explore paths with smaller probabilities thanks to these fluctuations (whereas in TDHF, only the most likely mean-field drives the dynamics). 

Experimentally, quasi-fission properties are often investigated by 
comparing reactions forming similar compound nuclei from different 
entrance channels \cite{chizhov2003,rafiei2008,williams2013}. 
A simultaneous investigation of quasi-fission and fission modes in the same system could be achieved by comparing 
$^{48,50}$Ca$+^{176}$Yb (beams of $^{50}$Ca  with sufficient intensity should be available at FRIB \cite{glasmacher2017}) in which one expects both quasi-fission and fusion-fission reactions, with $^{16,18}$O$+^{208}$Pb (forming the same compound nuclei) in which quasi-fission is expected to be negligible. 

 An experimental confirmation of strong similarities between (at least 
 some) fission and quasi-fission modes could help finding indications of
 the existence of theoretically predicted fission modes in nuclei that 
 are experimentally difficult to produce, such as in superheavy nuclei.
 Naturally, quasi-fission could not entirely replace the experimental 
 investigations of fission fragment distributions for the following 
 reasons: $(i)$ There is no guarantees that a mode observed in 
 quasi-fission would also be present in the fission of the compound 
 nucleus; $(ii)$ The relative abundance of competing quasi-fission modes 
 could be very different to that of fission modes; $(iii)$ Not all fission 
 modes are expected to be necessarily produced in quasi-fission.

%\section*{Declaration of Competing Interest}
%The authors declare that they have no known competing financial interests or personal relationships that could have appeared to influence the work reported in this paper.
%------------------------------------------------------------------------------
\section*{Acknowledgments}
We thank D. J. Hinde for his continuous support to this work, as well as R. Bernard for his help with the construction of the PES and a careful reading of the manuscript. 
Useful discussions with K. J. Cook and M. Dasgupta are also acknowledged. 
This work has been supported by the Australian Research Council Discovery Project (project number DP190100256) funding schemes, by the U.S. Department of Energy under grant Nos. DE-SC0013847 (Vanderbilt University) and DE-SC0013365 (Michigan State University), and by the NNSA Cooperative Agreement DE-NA0003885.
This work was also supported by computational resources provided by the Australian Government through the National Computational Infrastructure (NCI) under the ANU Merit Allocation Scheme.
%------------------------------------------------------------------------------
%------------------------------------------------------------------------------
\bibliographystyle{elsarticle-num}
%\bibliography{VU_bibtex_master.bib}

%\end{document}
\newpage
~~~
\newpage

%\appendix
%\section{SUPPLEMENTAL MATERIAL}

 \begin{table*}[hb!]
 {\bf Appendix. SUPPLEMENTAL MATERIAL}\\
 \setlength{\tabcolsep}{1.5pc}
 \newlength{\digitwidth} \settowidth{\digitwidth}{\rm 0}
\catcode`?=\active \def?{\kern\digitwidth}
 \caption{TDHF calculations for $^{50}\mbox{Ca}+^{176}\mbox{Yb}$ collisions at $E_{c.m.}=172$~MeV. The $^{176}$Yb deformation axis has an angle of 0, 45, 90 or 135 degrees with the line connecting the centre of masses of the nuclei in the TDHF initial condition. $\theta$ is the angle between the $^{176}$Yb velocity vector in the initial TDHF condition and its deformation axis.  $L$ is the initial orbital angular momentum of the collision. The contact time $\tau$ is given in zeptoseconds (1~zs$=10^{-21}$~s). The calculations are stopped when the contact time exceeds $\tau\sim31$~zs (unless the system is on its way to split in two fragments, in which case the calculation is run up to $\tau\sim35$~zs). The number of protons and neutrons in the heavy and light fragments are obtained from integration of the proton and neutron densities in the outgoing fragments. The total kinetic energy (TKE) is the sum of the kinetic energy of the fragments and of their Coulomb potential energy assuming point like fragments in the last TDHF iteration.\label{tab:results}}
 \begin{tabular*}{\textwidth}{lccccccc}
 \hline
 \hline
%$\beta$ (deg.) & $L$ ($\hbar$) & $\tau$ (zs)& $Z_H$ & $N_H$ & $\beta_{2_H}$ & $\beta_{3_H}$ & $Z_L$ & $N_L$ & $\beta_{2_L}$ & $\beta_{3_L}$& TKE (MeV)\\
$\theta$ (deg.) & $L$ ($\hbar$) & $\tau$ (zs)& $Z_H$ & $N_H$  & $Z_L$ & $N_L$ & TKE (MeV)\\
\hline
 14.7 &  70 &  \textgreater31.39 &        &        &        &        &        \\
 15.1 &  72 &  \textgreater31.39 &        &        &        &        &        \\
 15.5 &  74 &  \textgreater31.39 &        &        &        &        &        \\
 16.0 &  76 &  \textgreater31.39 &        &        &        &        &        \\
 16.4 &  78 &   28.37 &  53.48 &  81.82 &  36.52 &  54.18 & 164.50 \\
 16.8 &  80 &  \textgreater31.39 &        &        &        &        &        \\
 17.3 &  82 &   12.07 &  58.52 &  90.46 &  31.48 &  45.54 & 146.02 \\
 17.7 &  84 &   10.68 &  59.91 &  92.43 &  30.09 &  43.57 & 129.50 \\
 18.1 &  86 &   11.11 &  59.86 &  92.45 &  30.14 &  43.55 & 131.91 \\
 18.6 &  88 &    5.10 &  66.46 & 102.23 &  23.54 &  33.77 & 132.02 \\
\hline
 59.5 &  70 &  \textgreater31.22 &        &        &        &        &        \\
 59.9 &  72 &  \textgreater31.22 &        &        &        &        &        \\
 60.3 &  74 &   33.88 &  54.50 &  83.29 &  35.50 &  52.71 & 163.56 \\
 60.7 &  76 &   26.42 &  54.83 &  83.80 &  35.17 &  52.20 & 163.95 \\
 61.1 &  78 &  \textgreater31.23 &        &        &        &        &        \\
 61.6 &  80 &  \textgreater31.23 &        &        &        &        &        \\
 62.0 &  82 &   22.91 &  54.93 &  83.87 &  35.07 &  52.13 & 166.80 \\
 62.4 &  84 &   16.19 &  53.94 &  82.41 &  36.06 &  53.59 & 167.72 \\
 62.8 &  86 &   17.64 &  53.37 &  82.37 &  36.63 &  53.63 & 167.93 \\
 63.3 &  88 &   17.73 &  54.75 &  83.06 &  35.25 &  52.94 & 167.16 \\
 63.7 &  90 &   12.21 &  54.45 &  83.31 &  35.55 &  52.69 & 165.07 \\
 64.1 &  92 &   11.29 &  57.94 &  88.92 &  32.06 &  47.08 & 148.43 \\
 64.6 &  94 &    9.99 &  59.37 &  91.49 &  30.63 &  44.51 & 139.51 \\
 65.0 &  96 &    3.00 &  68.59 & 105.93 &  21.41 &  30.07 & 137.42 \\
\hline
100.5 &  50 &  \textgreater30.96 &        &        &        &        &        \\
100.9 &  52 &  \textgreater30.96 &        &        &        &        &        \\
101.3 &  54 &  \textgreater30.96 &        &        &        &        &        \\
101.7 &  56 &  \textgreater30.96 &        &        &        &        &        \\
102.1 &  58 &  \textgreater30.96 &        &        &        &        &        \\
102.6 &  60 &  \textgreater30.96 &        &        &        &        &        \\
103.0 &  62 &   13.33 &  56.90 &  87.23 &  33.10 &  48.77 & 148.65 \\
103.4 &  64 &   14.79 &  57.56 &  88.64 &  32.44 &  47.36 & 143.92 \\
103.8 &  66 &   12.85 &  56.58 &  86.97 &  33.42 &  49.03 & 143.38 \\
104.3 &  68 &   12.36 &  58.21 &  89.65 &  31.79 &  46.35 & 145.12 \\
104.7 &  70 &    4.90 &  65.88 & 100.66 &  24.12 &  35.34 & 132.51 \\
\hline
147.4 &  60 &  \textgreater31.17 &        &        &        &        &        \\
147.8 &  62 &  \textgreater31.17 &        &        &        &        &        \\
148.2 &  64 &  \textgreater31.17 &        &        &        &        &        \\
148.6 &  66 &   19.72 &  55.45 &  85.30 &  34.55 &  50.70 & 155.28 \\
149.0 &  68 &    2.81 &  68.55 & 105.29 &  21.45 &  30.71 & 135.83 \\
 \hline
 \hline
 \end{tabular*}
 \end{table*}

\end{document}